\documentstyle[twoside,fleqn,espcrc2]{article}

\newcommand{\twooneplaq}{\setlength{\unitlength}{.5cm}
   \raisebox{-.2cm}{
   \begin{picture}(2.2,1.2)(-1.1,-.6)
   \put(-1,-.5){\line(1,0){2}}
   \put(-1,.5){\line(1,0){2}}
   \put(-1,-.5){\line(0,1){1}}
   \put(1,-.5){\line(0,1){1}}
   \put(-1,-0.5){\circle*{.2}}
   \put(-1.5,-0.9){$x$}
   \put(-0.05,-1.2){$\mu$}
   \put(-1.5,-0.05){$\nu$}
   \end{picture}}}
\newcommand{\ltwooneplaq}{\setlength{\unitlength}{.5cm}
   \raisebox{-.2cm}{
   \begin{picture}(2.2,1.2)(-1.1,-.6)
   \put(-.5,-1){\line(1,0){1}}
   \put(-.5,1){\line(1,0){1}}
   \put(-.5,-1){\line(0,1){2}}
   \put(.5,-1){\line(0,1){2}}
   \put(-0.5,-1){\circle*{.2}}
   \put(-0.9,-1.5){$x$}
   \put(-1.2,-0.05){$\nu$}
   \put(-0.05,-1.5){$\mu$}
   \end{picture}}}

\newcommand{\AmS}{{\protect\the\textfont2
  A\kern-.1667em\lower.5ex\hbox{M}\kern-.125emS}}
\newcounter{arabiclistc}

\def\sqr#1#2#3{{\vcenter{\hrule height.#2pt
      \hbox{\vrule width.#2pt height#1pt \kern#3pt
         \vrule width.#2pt}
      \hrule height.#2pt}}}

\def\mystrut{{\vrule height 13pt depth 1.4pt width 0pt}}

\title{Topology by Improved Cooling: Susceptibility and Size Distributions
\thanks{
Talk by Margarita Garc{\'{\i}}a  P\'erez at LATTICE96.}}

\author{Philippe  de  Forcrand\address{SCSC, ETH-Z\"urich, CH-8092 Z\"urich, 
Switzerland}%
, Margarita  Garc{\'\i}a  P\'erez\address{Instituut Lorentz, 
Rijksuniversiteit Leiden, PO Box 9506, NL-2300 RA Leiden, Nederland}%
and Ion-Olimpiu  Stamatescu\address{FEST, Schmeilweg 5, D-69118 Heidelberg, 
Germany
\\  and \\
Inst. Theor. Physik,  Univ. Heidelberg, D-69120 Heidelberg, Germany
}}

\begin{document}

\begin{abstract}
We use a cooling  algorithm based on 
an improved action with scale invariant instanton solutions, which 
needs no monitoring
or calibration and has a inherent cut off for dislocations. In an
application to the $SU(2)$ theory
the method provides good susceptibility data
 and physical size distributions of instantons. 
\end{abstract}

\maketitle

\section{THE IMPROVED COOLING} 

In \cite{MNP} we presented an improved
cooling algorithm which fulfills requirements qualifying it as
 a method to eliminate UV noise and study topological 
excitations on the lattice:\par
\noindent - it smoothes out the short range fluctuations, 
including ``dislocations"\par
\noindent - it ensures  stability  
 of instantons (including their size)  above a certain threshold
 $\rho_0 \simeq 2.3a$  and needs thus no monitoring or calibration.\par
\noindent We apply this method to SU(2) to study susceptibility,
 size distribution, distances  and shape of instantons and 
 we report here results concerning scaling and boundary condition effects.

Since the Wilson action decreases with the instanton size, 
it leads to an unphysical
abundance of small instantons in a Monte Carlo
simulation. Under Wilson cooling  
even wide instantons shrink and decay.
This can be corrected for by modifying the lattice action.
Starting from  \cite{MP} one can construct a one parameter
set of actions with  no ${\cal O}(a^2)$ and ${\cal O}(a^4)$ corrections
using five fundamental, planar loops of size $m \times n$:
\begin{eqnarray}
S_{m,n}\hspace{-.3cm} &=& \hspace{-.2cm}\sum_{x,\mu,\nu} {\rm Tr} 
\left( 1 -\frac{1}{2}\ 
 ( \   \twooneplaq + \hspace{-.1cm}\ltwooneplaq\ \hspace{-.1cm} 
) \right) \\
S\hspace{-.3cm} &=& \hspace{-.2cm}  \sum_{i=1}^5 c_i \ 
{1 \over {m_i^2 n_i^2}}
\ S_{m_i,n_i}
\end{eqnarray}
\noindent Here  $(m_i,n_i) = (1,1),(2,2),(1,2),(1,3),(3,3)$
for $i=1,\ldots , 5$ and 
\begin{eqnarray}
c_1 &=&\hspace{-0.2cm}(19- 55\  c_5)/9,\ \  c_2 =  (1- 64\  c_5)/9 \nonumber\\
c_3 &=&\hspace{-0.2cm}(-64+ 640\  c_5)/45,\ \  c_4 = 1/5 - 2\  c_5
\end{eqnarray}
\noindent Our choice for an improved action is defined with the tuning 
$c_5=1/20$. The topological charge is improved the same way. 
We shall call this loop combination $5Li$: it will be exclusively used 
in the following.  As shown in \cite{MNP} it ensures 
scale invariance for instantons above a threshold $\rho_0 \simeq 2.3a$. 
The cooling algorithm exactly minimizes the local action at each 
step and involves no further calibration or engineering. 
 The size dependence of the action for the instantons
is $\theta$ -- function like,  giving  $8\pi^2$ to better than $0.1\%$
for $\rho > \rho_0$ and a steep descent (slightly configuration
dependent) in the interval $(0.8 - 1.0)\rho_0$. Since the threshold is
fixed in lattice units, it should not affect physical sizes if $a$ is 
small enough. For definiteness we use everywhere as radius in lattice
units:
\begin{equation}
\rho^4 \equiv \rho_{peak}^4 = 6 / \left( \pi^2 Q_{peak} \right)
\end{equation}
with $Q_{peak}$ the topological charge density at the center of the
instanton (using $S_{peak}$ instead gives no significant 
difference). Notice that the $Q(5Li)$
charge operator produces reasonable density data already on still rough
configurations and can be used to observe instanton-anti-instanton (IA) pairs
(which annihilate later in the cooling since they are not
minima of the action).

The question of stability of instantons cannot be completely disentangled from
finite size effects.
It is proven that $Q=\pm1$ solutions do not
exist on finite periodic volumes \cite{BP}. Practically this means that
isolated instantons above $\simeq 1/4$th of the lattice size are unstable.
To avoid this affecting the physical distances 
one can use twisted boundary 
conditions or make at least one of the lattice sizes very large
--- both these conditions allow to stabilize  wider instantons \cite{MP}.
The results of \cite{MNP} were obtained using  
twisted b.c. with $k=(1,1,1)$, $m=(0,0,0)$ (t.b.c. in the following).

The analysis undertaken here concerns the determination of 
physical quantities like 
topological susceptibility and size distribution 
in view of the two problems raised above, namely (A) threshold effects 
(for rough lattices
 $\rho_0$ may already represent a physically relevant distance) and
(B) finite size effects (small lattices  may not accommodate large, 
physical instantons). Correspondingly we analyze here
the scaling behavior and the dependence on size and boundary conditions.

\section{RESULTS AND DISCUSSION}

We base our analysis on the following 
$SU(2)$ Monte Carlo simulations (heat--bath, 
Wilson):\par
\noindent (1): $12^3\times N_t, \ \beta= 2.4$, ($a \simeq 0.12$fm) (a): $N_t=12$, t.b.c., 
(b): $N_t=12$, p.b.c., and (c): $N_t=36$, p.b.c.;\par
\noindent (2): $12^4, \ \beta=2.5$, ($a \simeq 0.085$fm), t.b.c.;\par
\noindent (3): $24^4, \ \beta=2.6$, ($a \simeq 0.06$fm), t.b.c..\par
\noindent Here (1) and (3) are new data from 200, respectively 
84  configurations (100, respectively 200
sweeps apart, after 20000 sweeps for thermalization), (2) are
older data \cite{MNP} from 160 configurations (250 sweeps apart).

{\it Topological Susceptibility:} Since the topological charge $Q(5Li)$ stabilizes very
fast to an integer within less than $1\%$ (see Fig. 1), we expect 
deviations from the physical values in the susceptibility to be only 
of type (A) or (B) above. The results are presented in Fig. 2 and Table 1.
The susceptibility settles early in the cooling and
 seems to scale very well both with the cut-off (compare the (1) and (3)
data) and with the 
volume (compare (1) and (2)), with small deviations 
showing up in (1) (beta=2.4) which will be argued to be mostly of
type (A). Assuming
the Witten -- Veneziano formula to hold the result agrees excellently with
the phenomenological expectation. Recent results for SU(2) in \cite{Ha}
 are about  $15\%$ higher than the ones we quote.  \par
\begin{figure}[htb]
\vspace{2.5cm}
\includegraphics{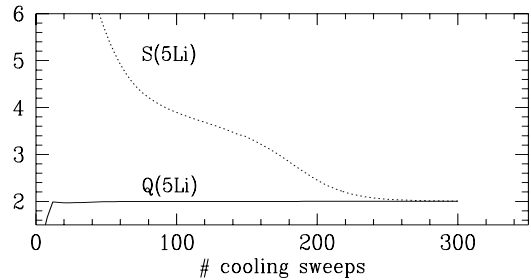}
\caption{Evolution of action and charge of a typical
configuration (t.b.c)  under $5Li$ cooling.}
\label{fig 1}
\end{figure}

\begin{figure}[htb]
\vspace{2.5cm}
\includegraphics{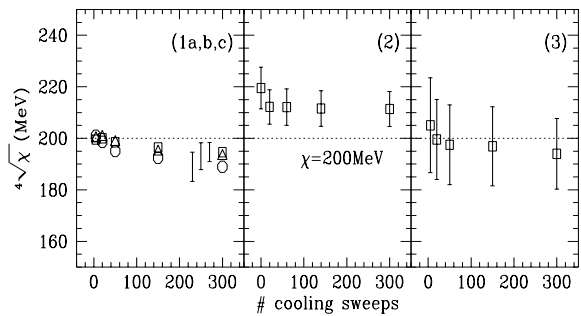}
\caption{Topological Susceptibility: t.b.c. (1a), (2) and (3) (squares),  
and  p.b.c. (1b) (triangles) and (1c) (circles) -- see text.
The bars in (1) give an indication of the errors which
are not included in the data to make the figure clearer.}
\label{fig 2}
\end{figure}

\vskip-0.5cm
\hbox to \hsize{\hfil\vbox{\offinterlineskip
\halign{&\vrule#&\ $#\mystrut$\hfil\ \cr
\noalign{\hrule}
&{\rm sw}&&(1{\rm a)}&&(1{\rm b)}&&(1{\rm c)} &&(3)&&{\rm sw}&&(2)&\cr
height 1.4pt&\omit&&\omit&&\omit&&\omit&&\omit&&\omit&&\omit&\cr
\noalign{\hrule}
height 1.4pt&\omit&&\omit&&\omit&&\omit&&\omit&&\omit&&\omit&\cr
&5&&199(3)&&200(5)&&201(5)&&205(16)&&0&&219(8) &\cr
&20&&200(3)&&200(6)&&199(5)&&199(15)&&20&&212(7)&\cr
&50&&198(3)&&198(6)&&195(5)&&197(15)&&60&&212(7)&\cr
&150&&196(4)&&195(6)&&192(6)&&197(15)&&140&&212(7)&\cr
&300&&195(4)&&193(6)&&189(6)&&194(14)&&300&&211(7)&\cr
\noalign{\hrule}}}\hfil}
\vskip3mm
{{\noindent Table 1: $^4\hspace{-1mm}\sqrt{\chi}$ (MeV) {\it vs} number of cooling sweeps.}\par}
\vskip5mm

{\it Instanton Size Distribution:} The size of the instantons is 
calculated following 
the continuum ansatz. We verified that a discretized form of this ansatz 
with periodicity corrections and generalized to an ellipsoid 
fits very well the $5Li$-densities 
(charge and action) of isolated instantons already after few 
cooling sweeps and that
it can also handle overlapping structures. 
The fitted radius averaged over directions agrees 
quite well with $\rho_{peak}$  Eq.(4).
Wide instantons above $1/2N_s$ show considerable anisotropy
and for such configurations different definitions of $\rho$ 
may lead to different values. We have restricted to the $\rho_{peak}$ definition
 except for  the old data in \cite{MNP} where the size
was extracted  differently. Consequently the upper part of the size 
distribution for (2) \cite{MNP} cannot be directly compared 
with (1) and (3).
\begin{figure}[htb]
\vspace{6.30cm}
\includegraphics{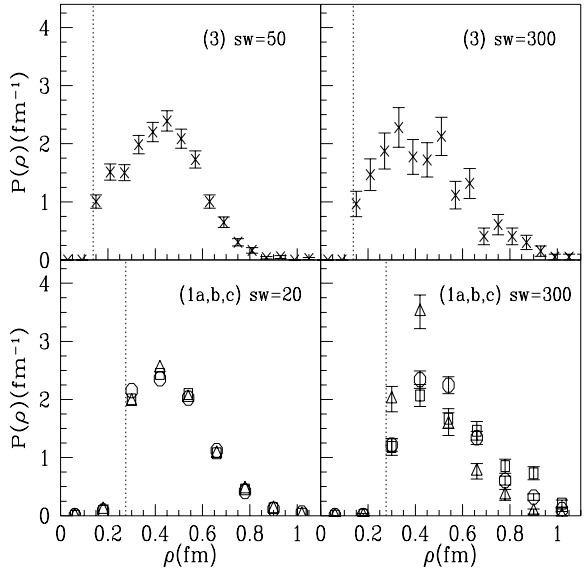}
\caption{Size distributions.
Crosses, squares, triangles, circles correspond respectively
to lattices (3), (1a), (1b), (1c), the dotted vertical line is $\rho_0$.}
\label{fig 3}
\end{figure}

The 
new results are presented in Fig. 3. The comparison of the (1) and (3) data
shows that at $\beta = 2.4$ the size distribution already begins to get
out of the threshold region, indicating a reduced effect of the latter
 on physical instantons.
The slight decay of the susceptibility observed in (1) can be completely 
understood as due to the disappearance of few, small but still physically
relevant instantons close to the threshold.

Finite size effects can be estimated by comparing (1a,b,c). The results
agree perfectly for low cooling sweeps but  for (1b) ($12^4$, p.b.c)
a shift towards smaller sizes at 300 sweeps is observed. We believe it to be 
mostly due to the instability of isolated instantons with p.b.c..
This conclusion is further supported  by comparing the
distributions (1a,b) and (1a',b') obtained from (1a,b) by
 leaving out configurations of $S=8\pi^2$ (see Fig. 4).\par

\begin{figure}[htb]
\vspace{3.3cm}
\includegraphics{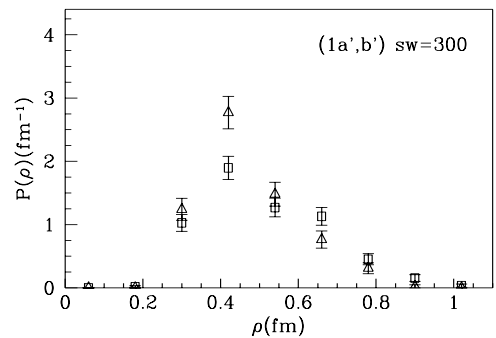}
\caption{Size distribution for (1a) and (1b) once subtracted 
the configurations with  $S=8\pi^2$.}
\label{fig 4}
\end{figure}

More detailed results, including larger statistics for $\beta=2.6$
and an analysis for quantities like distance distributions,
cross -- correlations of size and distances and charge distributions
will be presented elsewhere.

{\bf Acknowledgments}: FOM support for MPG and DFG support for IOS is thankfully acknowledged, likewise provision of computer facilities by the Universities of Heidelberg and Karlsruhe.

\end{document}